\documentclass[aps,prb,showpacs,twocolumn]{revtex4-1}

\usepackage{graphicx,bbm}

\def\tr{\,{\rm tr}\,}

\def\ave#1{\langle #1 \rangle}
\def\ii{{\rm i}}

\def\etal#1{#1}

\def\tit#1{}

\begin{document}

\title{Diffusive high-temperature transport in the one-dimensional Hubbard model}
 
\author{Toma\v z Prosen and Marko \v Znidari\v c}
\affiliation{Department of Physics, Faculty of Mathematics and Physics,
  University of Ljubljana, Ljubljana, Slovenia}

\date{\today}

\begin{abstract}
We consider charge and spin transport in the one-dimensional Hubbard model at infinite temperature, half-filling and zero magnetization. Implementing matrix-product-operator simulations of the non-equilibrium steady states of boundary-driven open Hubbard chains for up to $100$ sites we find clear evidence of diffusive transport for any (non-zero and finite) value of the interaction $U$.
\end{abstract}

 

\maketitle

\section{Introduction}

The one-dimensional (1d) Hubbard model is the simplest model of strongly interacting dynamics of spinful fermions on a lattice within a single-band approximation.
Its Hamiltonian reads
\begin{equation}
H=-t\sum_{i=1}^{L-1}\sum_{s\in\{\uparrow,\downarrow\}}(c_{i,s}^\dagger c_{i+1,s}+{\rm h.c.})+U\sum_{i=1}^L{n_{i\uparrow}n_{i\downarrow}},
\label{eq:hubbard}
\end{equation}
where the operators $c_{i,s}$, $c^\dagger_{i,s}$, for spin $s\in\{\uparrow,\downarrow\}$ and site $i\in\{1,\ldots,L\}$ are standard fermionic annihilation and creation operators and $n_{i,s}:=c^\dagger_{i,s} c_{i,s}$ are the density operators. The model is solvable by a coordinate Bethe ansatz~\cite{lieb} and possesses an infinite number of conservation laws~\cite{shastry:88}. While stationary properties of the 1d Hubbard model are well understood, see the monograph~\cite{book}, much less is known about its dynamics, for instance about the transport behavior. In thermodynamically the most interesting regime, at half-filled band and zero magnetization $\sum_{i=1}^L \ave{n_{i,s}} = L/2$, studied in the present paper, the model is gapped for charge excitations and gapless for spin excitations, for any $U\neq 0$. At zero temperature it is therefore an example of a Mott (charge) insulator and ballistic (ideal) spin conductor. Transport can be qualitatively characterized by a Drude weight -- a linear response indicator of ballistic transport, defined as the weight of zero-frequency singular term $\delta(\omega)$ in the real part of conductivity $\sigma(\omega)$. Spin and charge Drude weights at zero temperature have been calculated in Ref.~\cite{shastry:90}, with finite-size corrections given in Ref.~\cite{finiteL}, while a regular part of $\sigma(\omega)$ is studied in Ref.~\cite{finiteom}. At nonzero temperature on the other hand no rigorous result is known and there is no consensus between numerical and Bethe ansatz based results. Thermodynamic Bethe ansatz suggests \cite{fujimoto:98} that, even at half-filling, the charge Drude weight is finite, so the model was predicted to exhibit ideal charge transport; for a similar conclusion see also Quantum Monte Carlo calculation in Ref.~\cite{kirchner:99}. Analytical calculations at large $U$ on the other hand support vanishing charge Drude weight~\cite{peres:00}, for a study of low-energy excitations see also Ref.~\cite{gu:07}. Exact numerical simulations of small systems, again at half-filling and at high/infinite temperature, suggest~\cite{prelovsek}
$\sim 1/L$ scaling of the charge Drude weight. For temperatures much smaller than the gap semiclassical arguments together with field-theoretical scattering rate predicts diffusive transport~\cite{sachdev}. Vanishing finite-temperature Drude weight in thermodynamic limit (TL) $L\to\infty$ offers the possibility of an insulating or diffusive behavior.

However, at high or infinite temperatures, non-equilibrium transport properties of 1d Hubbard model in either charge or spin sector are not known as there has been up to date no analytical or numerical method capable of reliably treating this regime. In this paper we employ non-equilibrium steady state simulations~\cite{pz:09} using an efficient matrix product ansatz~\cite{vidal} for the time-dependent density matrix and provide a clear evidence of diffusive transport for both attractive $U<0$ and repulsive $U>0$ cases at infinite temperature. Namely, we show $1/L$ scaling of charge as well as of spin current and clear linear density profiles.

\section{Boundary driven Hubbard chain}

Using the Jordan-Wigner transformation we can map the 1d Hubbard model (\ref{eq:hubbard}) to a spin-$1/2$ ladder system. Namely, writing $c_{i\uparrow}=P^{(\sigma)}_{i-1} \sigma_i^-$
 where $P^{(\sigma)}_i=\sigma_1^{\rm z} \cdots \sigma_i^{\rm z}$ for spin-up fermions, 
 and $c_{i\downarrow}=P^{(\sigma)}_L P^{(\tau)}_{i-1} \tau_i^-$ 
 where $P^{(\tau)}_i=\tau_1^{\rm z} \cdots \tau_i^{\rm z}$ for spin-down fermions, one can verify that fermionic operators $c_{i,s},c^\dagger_{i,s}$ satisfy canonical anticommutation relations 
 provided $\sigma^\alpha_i$ and $\tau^\alpha_i$ are two sets of Pauli matrices (and $\sigma^\pm_j:=(\sigma^{\rm x}_j \pm\ii \sigma^{\rm y}_j)/2$, $\tau^\pm_j:=(\tau^{\rm x}_j \pm\ii \tau^{\rm y}_j)/2$).
 Writing the Hubbard Hamiltonian (\ref{eq:hubbard}) in spin-ladder form one obtains
\begin{eqnarray}
H=&-&\frac{t}{2}\sum_{i=1}^{L-1} (\sigma_i^{\rm x} \sigma_{i+1}^{\rm x}+\sigma_i^{\rm y} \sigma_{i+1}^{\rm y}+\tau_i^{\rm x} \tau_{i+1}^{\rm x}+\tau_i^{\rm y} \tau_{i+1}^{\rm y})+\nonumber \\
&+&\frac{U}{4}\sum_{i=1}^{L}(\sigma_i^{\rm z}+1)(\tau_i^{\rm z}+1).
\label{eq:ladder}
\end{eqnarray}
The spin-1/2 ladder system consists of two $XX$ chains in two legs and a $Z-Z$ type interchain coupling along the rungs. For numerical simulations of the Hubbard model we shall use this ladder formulation (\ref{eq:ladder}). 

To induce a nonequilibrium situation two legs are coupled to independent reservoirs. Their action is decribed in an effective way via the Lindblad equation~\cite{lin} for the density matrix $\rho$ of the ladder system,
\begin{equation}
\frac{{\rm d}}{{\rm d}t}{\rho}=\ii [ \rho,H ]+ {\cal L}^{\rm dis}(\rho),
\label{eq:Lindblad}
\end{equation}
where the dissipative term is expressed in terms of Lindblad operators $L_k$, as 
\begin{equation}
{\cal L}^{\rm dis}(\rho)=\sum_k \left( [ L_k \rho,L_k^\dagger ]+[ L_k,\rho L_k^{\dagger} ] \right).
\end{equation} 
We use eight Lindblad operators acting locally on the first and last sites of each leg, injecting or absorbing fermions (spinons) with certain probability:
\begin{equation}
L_{1,2}=\sqrt{\Gamma(1\mp \mu)}\,\sigma^{\pm}_1, \quad L_{3,4}=\sqrt{\Gamma(1\pm \mu)}\,\sigma^{\pm}_L,
\end{equation} 
for the first, and
\begin{equation}
L_{5,6}=\sqrt{\Gamma(1\mp \mu)}\,\tau^{\pm}_1, \quad  L_{7,8}=\sqrt{\Gamma(1\pm \mu)}\,\tau^{\pm}_L,
\label{eq:chargedriv}
\end{equation} 
for the second leg.
$\Gamma$ is the strength of the coupling to the baths while $\mu$ is a driving strength playing the role of a chemical potential bias. As demonstrated in previous studies of 1d spin chains~\cite{pz:09} the precise form of Lindblad operators does not influence the bulk properties. Because of dissipative terms the time-dependent solution $\rho(t)$ of the Lindblad equation converges after a long time to a time-independent state called a nonequilibrium steady state (NESS), $\rho_\infty = \lim_{t\to\infty}\rho(t)$, which is unique \cite{unique}.
Once the steady state  is reached, expressed in terms of a matrix product operator of a given bond dimension (following the method described in detail in Ref.~\cite{pz:09} straightforwardly adapted for the spin ladder), expectation values of arbitrary observables in the NESS can be efficiently evaluated. All expectation values considered in this paper are taken with respect to the NESS, that is $\langle A \rangle = \tr{(\rho_\infty A)}$, which we will -- when it is clear from the context, and to simplify notation -- denote just by $A$. In each NESS calculation we have carefully checked that the convergence is reached, i.e., we evolve the Lindblad equation (\ref{eq:Lindblad}) until a time-independent state is obtained, and that the results are stable with respect to increasing bond dimension \cite{technical}.  For $\mu=0$, i.e., no driving, one has an equilibrium setting, resulting in a trivial NESS $\rho_\infty \propto \mathbbm{1}$. This means that for small driving $\mu$ we are studying nonequilibrium behavior at an infinite temperature. Note that such an infinite temperature state is separable in the operator space. Since the efficiency of the numerical method crucially depends on the entanglement infinite-temperature nonequilibrium states are the easiest 
ones to calculate because the entanglement is expected to be smaller than at finite temperatures. 

Expectation values of fermionic observables are obtained from the corresponding ones in the ladder formulation, for instance, particle densities are $n_{i\uparrow}=(\sigma^{\rm z}_i+1)/2$ and $n_{i\downarrow}=(\tau^{\rm z}_i+1)/2$. Magnetization currents of the two spin species, defined through the continuity equations ${\rm d}(\sigma^{\rm z}_i/2)/{\rm d} t=j^{(\sigma)}_{i}-j^{(\sigma)}_{i-1}$, ${\rm d}(\tau^{\rm z}_i/2)/{\rm d} t=j^{(\tau)}_{i}-j^{(\tau)}_{i-1}$, are $j^{(\sigma)}_i=-\frac{t}{2}(\sigma_i^{\rm x} \sigma_{i+1}^{\rm y}-\sigma_i^{\rm y} \sigma_{i+1}^{\rm x})$, $j^{(\tau)}_i=-\frac{t}{2}(\tau_i^{\rm x} \tau_{i+1}^{\rm y}-\tau_i^{\rm y} \tau_{i+1}^{\rm x})$. In fermionic picture the particle (charge) current is $j^{(\rm c)}_i=j^{(\sigma)}_i+j^{(\tau)}_i$, while the spin current is $j^{(\rm S)}_i=(j^{(\sigma)}_i-j^{(\tau)}_i)/2$. Particle density is $n_i=n_{i\uparrow}+n_{i\downarrow}$, while spin density is $s_i=(n_{i\uparrow}-n_{i\downarrow})/2$.
Because of the same driving at both ladder legs the currents $j^{(\sigma)}$ and $j^{(\tau)}$ are the same. Therefore, NESS is such that it has a nonzero charge current and zero spin current. We have also performed simulations with Lindblad operators on the $\tau$-chain driving transport in the opposite direction, that  is with 
\begin{equation}
L_{5,6}=\sqrt{\Gamma(1\pm \mu)}\,\tau^{\pm}_1, \quad
L_{7,8}=\sqrt{\Gamma(1\mp \mu)}\,\tau^{\pm}_L.
\label{eq:spindriv}
\end{equation} 
In such a case of spin driving the NESS has a nonzero spin current and zero charge current because $j^{(\tau)}=-j^{(\sigma)}$ holds. Furthermore, we stress that spin and charge transport are interchanged under the particle-hole transformation for the {\em down} spin fermions only and simultaneously chaging the sign of $U$. Namely, taking $R:=\prod_{i=1}^L \tau^{\rm x}_i=R^\dagger$ one finds $R j^{({\rm S})}_i R^\dagger = j^{({\rm c})}_i/2$ and $R H(U) R^\dagger = H(-U)$, provided one takes a symmetric interaction term $(n_{i\uparrow}-\frac{1}{2})(n_{i\downarrow}-\frac{1}{2})$ in (\ref{eq:hubbard}) or, equivalently, adds a chemical term $-U N/2$ to $H$ with $N= \sum_{i,s}  n_{i,s}$. Even though our master equation evolution (\ref{eq:Lindblad}) does not strictly conserve $N$, we have checked explicitly that the results based on Hamiltonians $H$ and $H-U N/2$ are identical.

\section{Results}

\subsection{Evidence of diffusion: density profiles and scaling of currents}

\begin{figure}[h!]
\centerline{\includegraphics[width=0.47\textwidth]{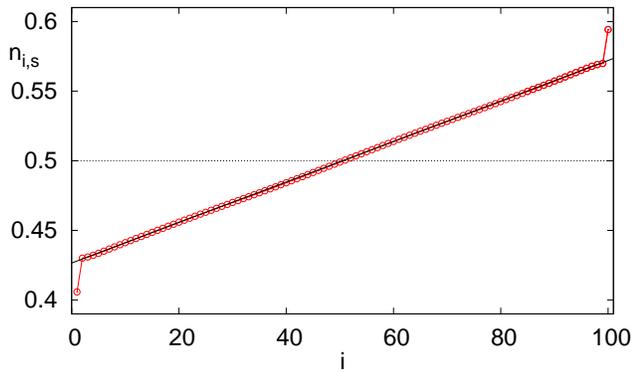}}
\caption{(Color online)~Density profile $n_{i,s}$ along the chain for $L=100$, $U=1$. Apart from jumps at the boundary, density is linear which is typical for diffusive conductors. Solid black line, overlapping with the numerical points, is a best-fitting linear function.}
\label{fig:profilU1}
\end{figure}

We set $t=1$, $\Gamma=1$ and $\mu=0.2$, except in Fig.~\ref{fig:profilmu1} where $\mu=1$. 
Driving $\mu=0.2$ corresponds to equilibrium density in the reservoirs of $n_{{\rm L},s}=0.4$ at the left end and $n_{{\rm R},s}=0.6$ at the right end. The average filling ratio is therefore $n=1/2$, $\sum_{i=1}^L n_{i\uparrow}=\sum_{i=1}^L n_{i\downarrow}=L/2$. The value $\mu=0.2$ is at the upper end of a linear response regime. For large drivings $\mu \gtrsim 0.6$ one gets a negative differential conductance effect \cite{benenti:09}, where the current decreases with increasing driving. The main goal of this paper is to classify spin and charge transport, whether it is {\em ballistic}, {\em diffusive} or {\em anomalous}. For NESS, different transport regimes are reflected in the scaling of the current on the system size. Fixing the driving strength $\mu$, in a ballistic conductor the current is independent of the system length, $j \sim L^0$, for a diffusive conductor it scales as $j \sim 1/L$, whereas in the anomalous case the current is proportional to a fractional power of $L$. We therefore calculated NESSs for different sizes $L$. Typical density profile is shown in Fig.~\ref{fig:profilU1}. 
One can see that the densities of spin-up and spin-down fermions are linear in the bulk. Jumps in the density at the boundaries are due to over-simplified Lindblad operators that are not ``matched'' to the bulk dynamics, i.e., there are boundary resistances. 
\begin{figure}[h!]
\centerline{\includegraphics[width=0.49\textwidth]{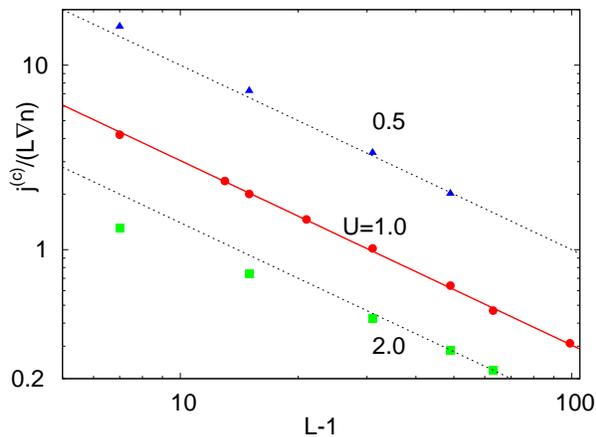}}
\caption{(Color online)~Scaling of charge current $j^{({\rm c})}$ divided by the extrapolated density drop $L\nabla n$ with the system size $L$ for different interactions $U$. Thick full (red) line, overlapping with $U=1.0$ data, is $\sim 30.4/L$, indicating a diffusive transport. Two dashed lines also suggest $\sim 1/L$ scaling.}
\label{fig:jodn}
\end{figure}
\begin{figure}[h!]
\includegraphics[width=0.53\textwidth,angle=-90]{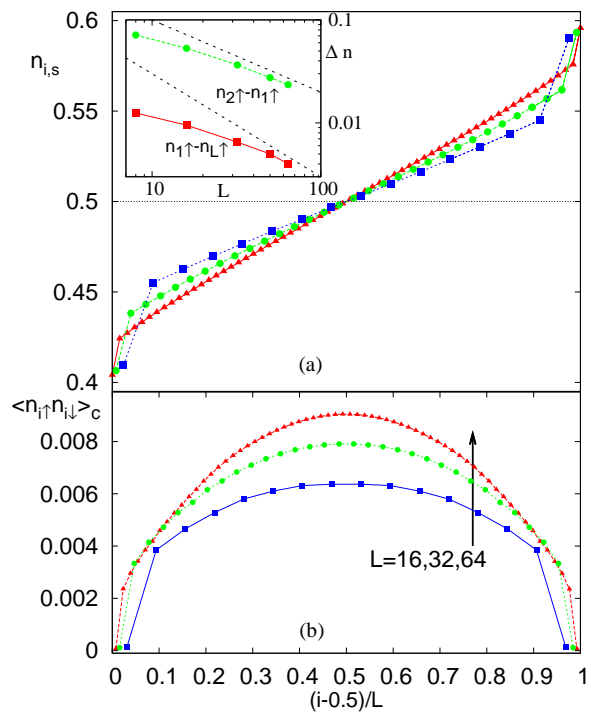}
\caption{(Color online)~(a) Density profiles $n_{i,s}$ at $U=2$, and (b) the corresponding connected density-density correlation function $\ave{ n_{i\uparrow} n_{i\downarrow}}_{\rm c}$. Data are shown for $L=16$ (blue squares), $L=32$ (green circles) and $L=64$ (red triangles), all for charge driving. Inset in (a): scaling of the jump between the reservoir and the 1st particle and between the 1st and 2nd particles, with size $L$. Black dashed lines indicate $\sim 1/L$ and $\sim 1/L^{0.7}$. Note that at $L=64$ boundary jumps still account for around $25\%$ of the total density difference between the chain ends.}
\label{fig:U2}
\end{figure}
Because these jumps are rather large we have fitted a linear function to the density profile in the bulk, thereby obtaining the density gradient $\nabla n_{i\uparrow}=\nabla n_{i\downarrow}=\nabla n_i/2$. In Fig.~\ref{fig:jodn} we then plot the scaling of the charge current (which is in the NESS independent of the site) with the gradient of the charge density. At interaction strength $U=1$ one can see a nice scaling $j^{({\rm c)}} \sim 1/L$. Together with a linear density profiles this is a clear indication of diffusive charge transport. As mentioned, for spin transport virtually the same behavior is obtained (data not shown). 
For $U=2$ the scaling is not quite as good. It seems that for shorter chains $j^{({\rm c})}$ decreases with $L$ slower than $1/L$, however, for two largest sizes that we managed to calculate, a crossover to $\sim 1/L$ scaling is clearly suggested. For smaller interaction $U=0.5$ the convergence seems better, but contrary to the $U=2$ case, the current approaches the asymptotic scaling ($\sim 1/L$) from above, i.e. it decays a bit faster for short chains. It is not clear whether $U=1$ corresponds to a crossover point between the two behaviors since details of convergence in the thermodynamic limit might depend on a particular choice of boundary Lindblad operators. In Fig.~\ref{fig:profilU0.5all}(a) we show density profiles for $U=0.5$ which, apart from considerable boundary jumps, again look linear.

\begin{figure}[h!]
\centerline{\includegraphics[width=0.47\textwidth,angle=-90]{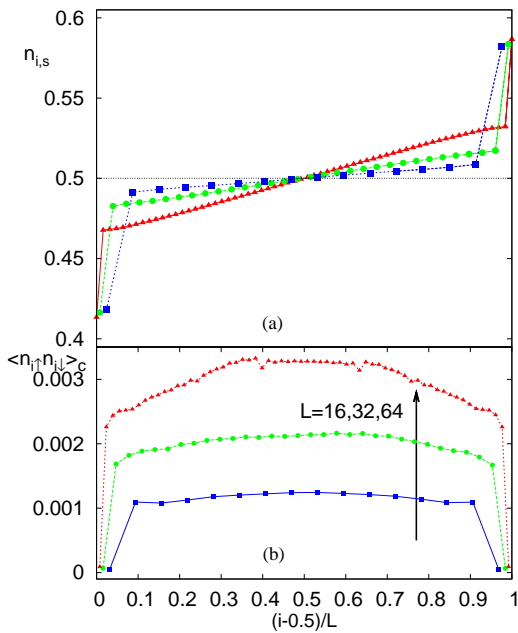}}
\caption{(Color online)~Density profiles (a) and connected correlations (b) for $U=0.5$. Other parameters are the same as in Fig.~\ref{fig:U2}.}
\label{fig:profilU0.5all}
\end{figure}

Looking at the density profiles at $U=0.5,1,2$ which are linear in the bulk already for rather small sizes $L$, it seems natural to conjecture that the transport is diffusive in TL $L \to \infty$ for all nonzero finite values of $U$. Different transient scaling of the current with $L$ for shorter chains is likely due to rather strong boundary effects. That the boundary effects are notable can also be seen in the inset of Fig.~\ref{fig:U2}(a), where we show the jump in the density between the reservoir and the first site $n_{1\uparrow}-n_{{\rm L}\uparrow}$, as well as between the first two sites in the system $n_{2\uparrow}-n_{1\uparrow}$. While the boundary effects show a tendency to disappear in TL, at $U=2$ and largest length $L=64$ they are still non-negligible. One can try to optimize the coupling constant $\Gamma$ in order to minimize the boundary effects, however we found that $\Gamma \approx 1$ is usually close to the optimal value which does not seem to depend on $L$.
For $U=0.5$ the boundary effects are larger than for larger $U$ cases studied, $U=1$ and $U=2$, see Fig.~\ref{fig:profilU0.5all}(a). 
This is probably due to a smaller bulk resistivity which makes the effects of the contacts (contact resistivity) relatively larger.

\subsection{Density-density correlations}

\begin{figure}[h!]
\centerline{\includegraphics[width=0.47\textwidth,angle=-90]{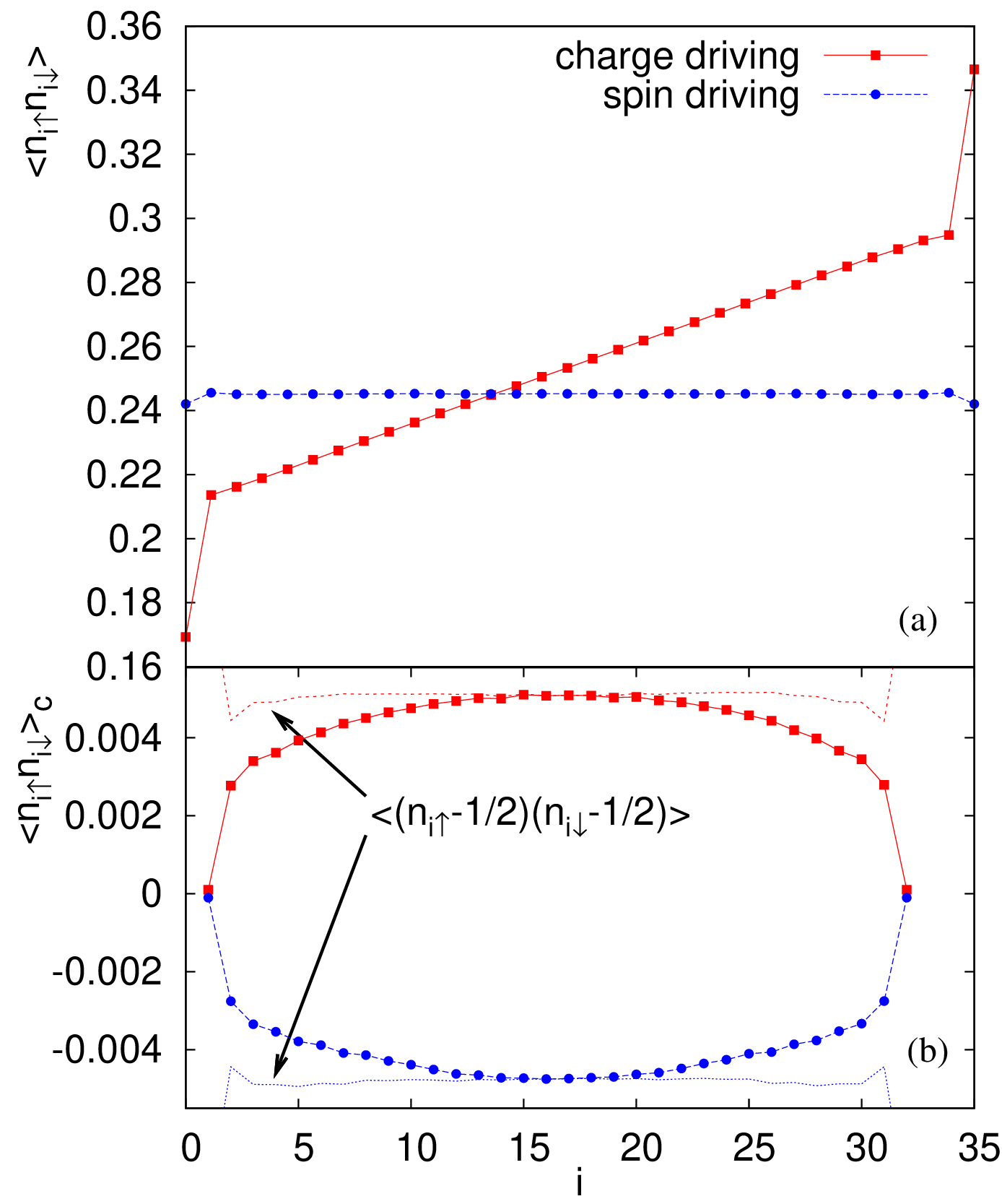}}
\caption{(Color online)~Density-density correlation functions for the case of charge driving (red squares) and for spin driving (blue circles). Frame (b) shows the connected correlations (symbols), frame (a) the non-connected ones. In (b) we also show $\langle (n_{i\uparrow}-\frac{1}{2})(n_{i\downarrow}-\frac{1}{2}) \rangle$ (dashed curves). All is for $U=1$, $L=32$.}
\label{fig:corelU1}
\end{figure}

In Figs.~\ref{fig:U2}(b),~\ref{fig:profilU0.5all}(b), and also ~\ref{fig:corelU1}(b), we show the connected spin-up spin-down correlation function $\langle n_{i\uparrow}n_{i\downarrow}\rangle_{\rm c}=\langle n_{i\uparrow}n_{i\downarrow} \rangle - \langle n_{i\uparrow} \rangle \langle n_{i\downarrow} \rangle$, that gives on-site correlations between two fermion species. If we extrapolate our finite-$L$ data to TL, we find \begin{equation}
\langle n_{i\uparrow}n_{i\downarrow}\rangle_{\rm c} \propto (\mu^2-(2\ave{n_{i\uparrow}}-1)(2\ave{n_{i\downarrow}}-1)), 
\end{equation}
with a proportionality prefactor depending on $U$ only, yielding a {\em parabolic correlation profile} for our {\em linear} density profiles. Interestingly, in the middle of the chain the connected correlations become independent of the system size, while they are going to zero at the boundaries\cite{footnotea}. 
In Fig.~\ref{fig:corelU1} we in addition show also, for $U=1$ case, the non-connected correlations $\ave{n_{i\uparrow}n_{i\downarrow}}$ (top frame (a), red squares), or centered non-connected correlations $\ave{(n_{i\uparrow}-1/2)(n_{i\downarrow}-1/2)}$ (bottom frame (b), red dotted line). We can see that the non-connected correlations have similar shapes as density profiles.

Considering spin driving (\ref{eq:spindriv}), we find numerically the same diffusion constant as for charge driving (\ref{eq:chargedriv}), while the connected correlations $\langle n_{i\uparrow}n_{i\downarrow}\rangle_{\rm c} $ change sign (see Fig.~\ref{fig:corelU1}(b), blue circles). Note that this implies that conductivities at infinite temperature do not depend on the sign of interaction $U$. 
We have also checked explicitly, by comparing data for $U=-1$ with $U=1$, that  density profiles, currents and correlations are insensitive to the sign of $U$.
One can see that in the presence of spin current without charge current, non-connected correlations, shown in Fig.~\ref{fig:corelU1}(a) with blue circles, are practically independent of the site and are slightly smaller than $1/4$. 

Note that in both cases, of charge and spin driving, the connected correlations scale as $\sim \mu^2$ and are of purely nonequilibrium origin, i.e. they vanish in the equilibrium limit ($\mu=0$).

\subsection{Large interaction $U$}

In the limit $U \to \infty$ the low energy excitations of the half-filled Hubbard model can be effectively described by the 1d isotropic Heisenberg model. In our open system formulation this
mapping cannot be strictly implemented, due to the presence of high-temperature baths which drive the system locally away from half-filling. It is therefore an interesting question whether the transport properties of the Hubbard model in the limit of large $U$ are qualitatively the same as for the Heisenberg model.

\begin{figure}[h!]
\centerline{\includegraphics[width=0.47\textwidth]{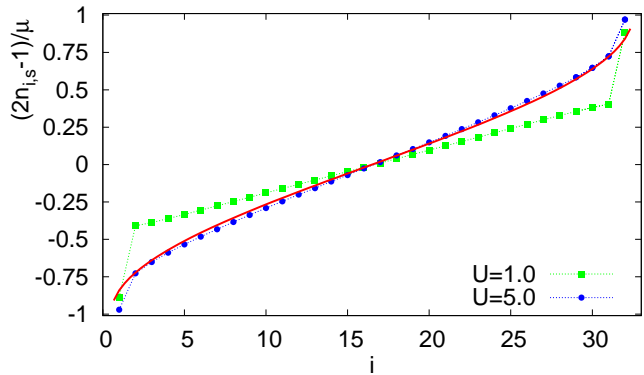}}
\caption{(Color online)~As one increases $U$ at fixed length $L=32$ the rescaled density profiles $(2n_{i,s}-1)/\mu$, shown for $U=1$ and $U=5$, become similar to $\frac{2}{\pi}\arcsin{[\frac{2i-1}{L}-1]}$ (full red curve), found in the isotropic Heisenberg model \cite{znidaric:11}.}
\label{fig:fixedn}
\end{figure}

In Fig.~\ref{fig:fixedn} we plot density profiles in our open Hubbard model for increasing values of $U$, keeping $L$ fixed, and find
increasingly $\arcsin{}$-like shape, similar as in the isotropic Heisenberg model which displays an anomalous transport~\cite{znidaric:11} with the magnetization current scaling as $\sim 1/\sqrt{L}$.
This perhaps explains slower decay of the current with $L$ in the Hubbard model for small $L$'s and larger $U$, seen for instance in Fig.~\ref{fig:jodn} at $U=2$. Note that the limits $U \to \infty$ and $L \to \infty$ do not commute. In order to recover the Heisenberg behavior in TL one has to first let $U \to \infty$ and only then $L \to \infty$.

\subsection{Strong driving, $\mu=1$}

\begin{figure}[t!]
\centerline{\includegraphics[width=0.47\textwidth]{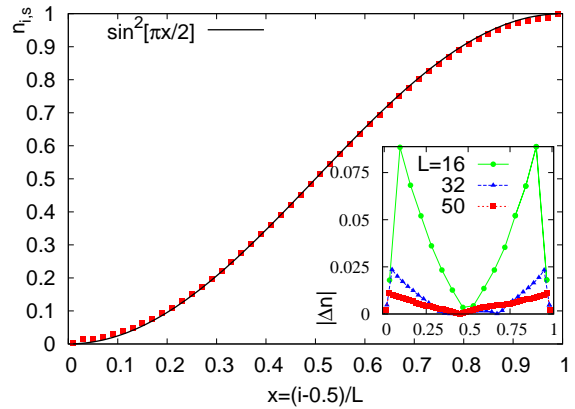}}
\caption{(Color online)~At maximal driving, $\mu=1$, density profiles have a cosine shape (red squares, for $L=50$, $U=1$) while the current scales as $\sim 1/L^2$ (data not shown), exactly as in the Heisenberg model at maximal driving \cite{prosen:11}. In the inset we show convergence of $\Delta n=n_{i,s}-\sin^2{(\pi x/2)}$ with $L$.}
\label{fig:profilmu1}
\end{figure}

In previous subsections we have shown that the weakly driven Hubbard model displays diffusive behavior. Here we show that for strong driving, where the system is far away from equilibrium, the behavior
of physical observables can be dramatically different. For example, we briefly discuss the case of maximal driving $\mu=1$ and find that the current scales sub-diffusively as $j^{\rm c} \propto 1/L^2$.
The corresponding density profile is shown in Fig.~\ref{fig:profilmu1}. We can see that the profile is in TL given by a simple cosine shape $n_{i,s} = \sin^2(\pi (2i-1)/4L)$, exactly the same as has been found analytically in the isotropic Heisenberg model at strong driving~\cite{prosen:11}. This suggests that
similar exact solution for NESS at maximum driving $\mu=1$ as for the Heisenberg spin chain is also achievable for the Hubbard model and points to wider applicability of the algebraic method proposed in 
Refs.~\cite{prosen:11,prosen:11b}.

\begin{figure}[h!]
\centerline{\includegraphics[width=0.47\textwidth,angle=-90]{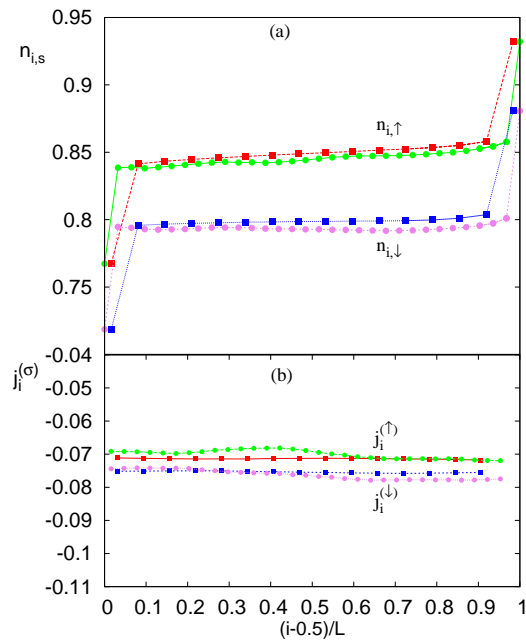}}
\caption{(Color online)~Density profiles (a) and currents (b) for a non-half-filled Hubbard system. We show data for $L=16$ (squares) and $L=32$ (circles), all for $U=1$. Density profiles in the bulk are flat and currents do not depend on $L$, indicating ballistic transport. In both frames upper two sets of symbols are for one fermion species, lower two for the other.}
\label{fig:nonhalf}
\end{figure}

\subsection{Non-half-filled case}

So far, all the results shown were for symmetric driving, producing on average half-filled bands. However, if the filling of the two fermion species is not $1/2$ we expect the transport to be ballistic at high temperatures. This follows from an existence of a nontrivial constant of motion, which in the non-half filled case possesses nonvanishing overlap with the spin/charge currents~\cite{Zotos:97}. In order to numerically verify the consistency of our non-equilibrium setup with this expectation, we choose a nonsymmetric driving with $L_{1,2}=\sqrt{\Gamma(1\mp \mu_{\rm L\uparrow})}\,\sigma^{\pm}_1$ and $L_{3,4}=\sqrt{\Gamma(1\pm \mu_{\rm R \uparrow})}\,\sigma^{\pm}_L$ for the first chain, where $\mu_{\rm L\uparrow}=0.5$ and $\mu_{\rm R\uparrow}=0.9$, while $L_{5,6}=\sqrt{\Gamma(1\mp \mu_{\rm L \downarrow})}\,\tau^{\pm}_1$ and $L_{7,8}=\sqrt{\Gamma(1\pm \mu_{\rm R\downarrow})}\,\tau^{\pm}_L$ with $\mu_{\rm L\downarrow}=0.4$ and $\mu_{\rm R\downarrow}=0.8$ for the 2nd chain. Density profiles can be seen in Fig.~\ref{fig:nonhalf}. One can see that the gradient is very small (or zero); what is more, currents are almost independent of system size $L$. Namely, in Fig.~\ref{fig:nonhalf}b currents are within numerical errors the same for $L=16$ and $L=32$ (small inhomogeneities visible in the Figure are due to truncation errors). If the transport were diffusive, as is the case for the half-filled system, the current for $L=32$ should be half as large as for $L=16$. We therefore confirm that the non-equilibrium transport is clearly ballistic for a non-half-filled Hubbard model.

\section{Conclusion}

Summarizing our findings about the transport in half-filled zero-magnetization 1d Hubbard model at an infinite temperature, we have shown that at finite interaction $U$ both charge and spin transport are diffusive. This conclusion is based on the scaling of the currents with the system size for up to $100$ particles, as well as on perfectly linear density profiles away from the boundaries. This complements ballistic spin transport at $U=0$ and arbitrary temperature and anomalous transport at high temperature and $U=\infty$. We acknowledge support by the grants P1-0044 and J1-2208 of Slovenian Research Agency (ARRS).


\begin{thebibliography}{1}
\bibitem{unique} NESS is in our case unique and is therefore reached irrespective of the initial state $\rho(0)$. Evans theorem \cite{evans} states that NESS (fixed point of a Liouvillian dynamical semi-group (\ref{eq:Lindblad})) is unique iff the set of operators $\{H,L_1,L_2,\ldots\}$,
in our case $\{H,\sigma^\pm_{1,L},\tau^\pm_{1,L}\}$, generates, under multiplication and addition, the entire algebra of operators, in our case over the spin ladder on $L$ sites.
Indeed, this can be verified by operator identities giving a recursive generation of $\sigma^\pm_j$: 
$\sigma^+_2  = (1/4) \sigma^{\rm z}_1 [\sigma^+_1,[H,\sigma^{\rm z}_1]]$, and for $j \ge 2$,  $\sigma^+_{j+1} = -\sigma^+_{j-1}  - (1/2)\sigma^{\rm z}_j [\sigma^-_j,\sigma^+_j H \sigma^+_j] $.
Operators for the second leg $\tau^\pm_j$ are generated through a similar recursion, replacing $\sigma^\alpha_j$ by $\tau^\alpha_j$.
Finally, all possible products of $\{ \sigma^\pm_j \}$ and $\{ \tau^\pm_j \}$ generate the entire operator algebra of the Hubbard chain.
\bibitem{technical} We use bond dimensions up to $D=80$. Convergence times $t_\infty$ typically scale $\propto L$, for example for $L=100$ we needed to simulate (\ref{eq:Lindblad}) up to $t_\infty\approx 500$.
\bibitem{footnotea} Note that small noisy irregularities in the parabolic correlation profile in Fig.~\ref{fig:profilU0.5all}(b) for $L=64$, being of order $10^{-4}$, are due to insufficient matrix product bond dimension $D$. Nevertheless, we have checked that the steady state current, which is here $j^{\rm c} \approx 0.05$, has already converged with respect to increasing $D$.
\end{thebibliography}
\end{document}